\documentclass[final,1p,times]{elsarticle}
\usepackage{txfonts}
\usepackage{graphicx}
\usepackage{epsfig}
\usepackage{graphics}
\usepackage{amssymb}
\usepackage{subfig}
 \usepackage{amsthm}
\usepackage{amscd}
\usepackage{amsmath}
\usepackage{amsfonts}
\usepackage{amssymb}
\usepackage{graphicx}

\numberwithin{equation}{section}

\newcommand{\be}{\begin{equation}}

\newcommand{\ee}{\end{equation}}
\newcommand{\bea}{\begin{eqnarray}}
\newcommand{\eea}{\end{eqnarray}}
\newcommand{\bna}{\begin{eqnarray*}}
\newcommand{\ena}{\end{eqnarray*}}

\journal{$\ast\ast\ast$}

\begin{document}

\begin{frontmatter}

\title{Analysis on Cohort Effects in view of Differential Geometry and its Applications}
\author[label1]{Ning Zhang}
 \ead[label1]{nzhang@amss.ac.cn}
\address[label1]{China Institute for Actuarial Science,
 Central University of Finance and Economics, Beijing 100081, P. R. China}
\author[label2,label3]{Liang Zhao}
 \ead[label2]{ liangzhao@bnu.edu.cn}
\address[label2]{Corresponding author, School of Mathematical Sciences,
 Beijing Normal
University, Beijing 100875, P. R. China}
\address[label3]{Laboratory of Mathematics and
Complex Systems, Ministry of Education,
Beijing 100875, P. R. China}

\begin{abstract}
This paper analyzes birth cohort effects and develops an approach which is based on differential geometry to identify and measure cohort effects in mortality data sets.
The measurement is quantitative and provides a potential method to compare cohort effects among different countries or groups. Data sets of four countries (e.g. U.k., U.S., Canada and Japan) are taken as examples to explain our approach and applications of the measurement to a modified Lee-Carter model are analyzed.

\end{abstract}

\begin{keyword}
Mortality, Cohort Effect, Lee-Carter model, Normal Vector, Normal Curvature
\end{keyword}

\end{frontmatter}

\section{Introduction}

      Cohort effects are used in the social sciences to describe variations among individuals who are defined by some shared temporal experiences or common life experiences, such as year of birth, or year of exposure to radiation. Many papers and reports have considered cohort effects in mortality data and highlighted the existence of cohort effects in different countries \cite{cmi,novaneetham,willets}. For example, it is well known that people born in the U.K. between 1925 and 1945 have experienced more rapid improvement in mortality than generations born in other periods \cite{government}. In other words, this generation has experienced more obvious cohort effect than others. There is some further research on empirical analysis about the influence of cohort effects on mortality \cite{novaneetham,willets} and many conceptual advances in the social sciences have been generated through the idea that cohort effect is an important index that affects health and development through the life course \cite{alwin, schaie}. It is nature to ask how much is the cohort effect for a specific group and how to compare cohort effects between different groups? The attempt to evaluate cohort effects can be find in the paper by W. Farr in 1885 \cite{farr}. There are also some graphical approaches throughout the first half of the 20th century \cite{frost, kermack, susser}. More recently, statistical approaches to effect quantification are developed. But conventional statistical models such as generalized linear regressions cannot provide valid estimates of age, period and cohort effects directly because of colinearity of the three variables. Although there are some methods to overcome the obstacle, for example, one can refer to \cite{lee2,renshaw2,renshaw1} and the references therein, the lack of an accurate description of cohort effects is still a limit of the application of cohort analysis to increasing data sources. This is the main motivation of our paper.

      Since we can view a mortality data set as a discrete surface of dimension $2$, we can deal with this object geometrically. The more apparent cohort effect for some group is, the more fluctuant the surface corresponding to this group is. An appropriate concept in differential geometry to describe the fluctuate of a surface is  curvature. Our idea in this paper is to compute the curvature for certain group of people so that we can quantify the corresponding cohort effect. We use our algorithm to analyze the data set of several countries. The results are coincident with the literature and can even give some improvement.

      The enormous improvement in life expectancy is certainly one of the greatest achievements of modern civilization, but unanticipated mortality improvement can pose huge problems to individuals, corporations and governments. In fact, longevity risk, that is, the uncertainty associated with future mortality improvement, has become a prominent risk in many countries and the financial burden to individuals, corporations and governments is also triggered by mortality improvement. To model and analyze longevity risk, the cohort effect has become an important factor. As applications, we also apply our quantitative cohort effects in mortality models of longevity risk.

      Many models have been promoted to acquire information about longevity risk from mortality data sets. Among these models, the Lee-Carter model is widely used and its derivative models have shown their success in forecasting future mortality.
      In the Lee-Carter model, mortality can vary across individuals as they age (aging or life cycle effects) and time goes (period effects). Let $\mu_x(t)$ denotes the central death rate for age x at time t, the model can be expressed as the following log-bilinear form \cite{lee1}:
      $$ \ln \mu_x(t)=\alpha_x +\beta_x k_t +\epsilon_{x,t}.$$
      Here the change in the level of mortality over time is described by the mortality index $k_t$, $\alpha_x$ describes the age-pattern of mortality averaged over time and $\beta_x$ describes the deviation from the averaged pattern when $k_t$ varies. Finally, the quantity $\epsilon_{x,t}$ denotes the error term, with mean $0$ and variance $\sigma^2$. To make the model perfect, possible improvements of the Lee-Carter model are promoted, which use a random variable $D_x(t)$ for the number of deaths \cite{brouhns2}. Here $D_x(t)\sim {\text Poisson}(ETR_x(t),\mu_x(t))$ , $ETR_x(t)$ is the central number of exposed to risk which refers to the total number of person-years in a population over a calendar year in the actuarial literature and $\mu_x(t)$ is given by:
      $$\ln\mu_x(t)=\alpha _x +\beta_x k_t.$$

   The Lee-Carter model generally considers only age effects and period effects. But a group born in the same year should share some characters which have an impact on its mortality. To reflect this impact, we had better take cohort effects into consideration in the model. In fact, there are many works about the age-period-cohort analysis of mortality. We call this kind of models the APC model. For example, we can consider a simple model which includes cohort effects like this form \cite{willets}:
    \begin{equation*}\label{apc}
    \ln\mu_x (t)=m+\alpha_x+\beta_t+\gamma_{t-x}.
    \end{equation*}
    Here $\alpha_x$ denotes age effects, $\beta_t$ denotes period effects and $\gamma_{t-x}$ denotes cohort effects. There are also works by Renshaw and Haberman \cite{renshaw2,renshaw1} which introduce cohort effects into the Lee-Carter model. Their model is the following age-period-cohort (APC) version of the Lee-Carter model :
    $$ \ln \mu_x (t)=\alpha _x +\beta_x^{(0)}i_{t-x}+\beta_x^{(1)}k_t.$$
    Here an extra item $\beta_x^{(0)}i_{t-x}$ is introduced in order to represent cohort effects.  The model can give rise to two sub-structure models: setting $\beta_x^{(0)}=0$ to get the classical Lee-Carter model and setting $\beta_x^{(1)}=0$ to get an age-cohort model. When proceed the APC models, there is something may be troublesome. It is obvious that there exists a linear dependency of age (years since birth), period (year), and cohort (year of birth): Age$=$Period$-$Cohort and only two of these effects can be identified (introduction of this issue, see \cite{winship}). In other words, we can find no unique set of parameters resulting in an optimal fit because of this trivial relation. A widely used technique to deal with this problem is to add further constraints on parameters \cite{kupper,renshaw2}. Based on different constraints, there are widespread applications \cite{clayton,leelin,luostarinen} which can fit the data sets well. However, since there is not a clear explainable reason for the further constrains either from the point of view of mortality or from mathematics, it is possible that additional bias will be induced into the models.

    As an application of our estimates on cohort effects, we want to work under the framework of the APC model, which means that we still assume that the logarithm of the expected rate of mortality is a linear function of the three effects. Now since we have quantitative estimates for cohort effects, it is possible to get rid of this kind of effects in the APC model at first. Then the model will be converted into the framework of the Lee-Carter model and can be analyzed as a Lee-Carter model. We apply this idea to analyze mortality data of U.K., U.S., Canada and Japan and compare our results with the classical Lee-Carter model. It is found that our results are similar to the results from the Lee-Carter model. We also forecast mortality of U.S. by using both our method and the APC model. The results still indicate that these two methods are similar. Moreover, sometimes our predicted value is almost the same with the actual value.

    Now let us make a summary about what inspires us to find a way to detect and measure cohort effects:

     1) The birth cohort effects exist in some mortality data sets. Although we can observe a data set directly to find some evidences, we do not know the exact information about cohort effects when given any mortality data set.

     2) The cohort effect should be added into the Lee-Carter model. But this may bring some troubles in the calibration of the model.

     3) Quantitative estimates of cohort effects can help us to make population and health policy, to understand trends of epidemic and to design insurance product.

     The consequences of the paper include the followings:

     1) Get exact information about cohort effects for any mortality data set, which is an important population character for different groups.

     2) Apply our quantitative estimates of cohort effects to modify the Lee-Carter model and compare our results with results from the classical Lee-Carter model and the APC model.

     We organize the paper as follows: In Section 2, we propose the approach based on the theory of differential geometry, which attempts to identify cohort effects of any mortality data set. We also give a definition of cohort effect index(CEI) which can illustrate the strength of cohort effect in any period. In Section 3 we apply our approach to the mortality data sets of four countries and give some explanation of the results. In Section 4, we apply the quantitative cohort effect to modify the Lee-Carter model. We also give a prediction of mortality by using the modified Lee-Carter model and compare our forecasting results with the results of the APC model. Section 5 outlines some potential applications and further developments of this paper.

\section{Theoretical model of detecting cohort effects}

We use some fundamental concepts in differential geometry in this section which can be found in any textbook of differential geometry, for example one can refer to \cite{carmo, gray}. We start from an ideal situation that the mortality data set corresponds to a smooth surface $\Sigma_s$ which is called a smooth mortality surface and is defined by a smooth function $z=f(t,x)$, where $t$ denotes time, $x$ denotes age and $z$ is the mortality of the people at the age of $x$ when time is $t$.

The mortality data for people born in year $t$ represent a curve $l_t$ on $\Sigma_s$ and we call the curve a cohort curve. If cohort effects appear for these people, the mortality of this group should have general characters and should be distinguished from the mortality data of groups nearby. For this reason, at any $p\in l_t$, the oscillations along any other directions should be more observable than along the tangent vector of $l_t$. In differential geometry, an effective tool to measure this kind of oscillation is to compute the normal curvature of a curve on a surface. A cohort curve with cohort effects should be a curvature line and the tangent vector of this curve should be one of the principle directions (Figure 1).

\begin{figure}[tbh]
\begin{centering}
\includegraphics[height=150pt,width=330pt]{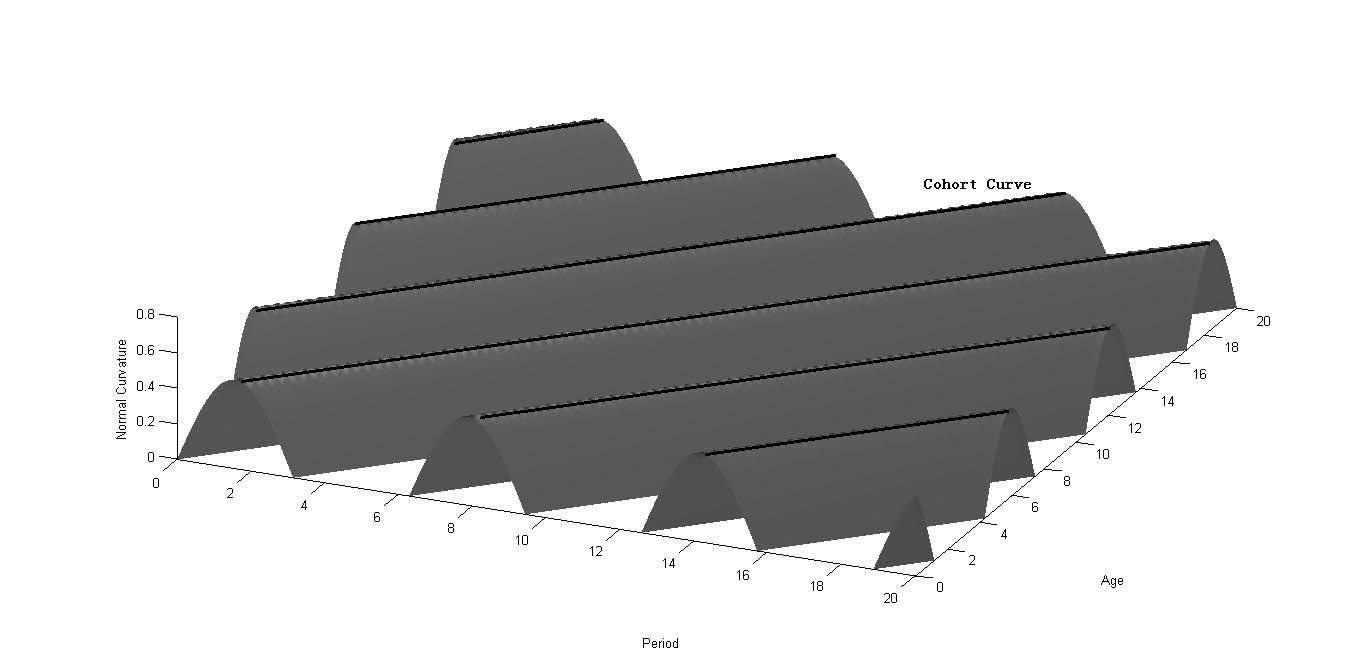}
\end{centering}
\caption[uk]{The sample of cohort curve which is not real,  but just for explanation}
\end{figure}

But, in fact, since there are many kinds of noises in mortality data and the data are discrete, we cannot expect a cohort curve with cohort effects to be identical to a curvature line. For this reason, we define the cohort effect index mathematically as follow.\\

\noindent{\bf Definition:} {\it For a cohort curve on $\Sigma_s$, suppose $T(s)$ is the tangent vector of a point $p(s)\in l_t$ and $N(s)$ is the orthogonal direction of $T(s)$ on the tangent plane of $p(s)$. Let $NC_T (s)$ and $NC_N (s)$ be the normal curvatures along the directions $T(s)$ and $N(s)$ respectively. Then the integral
\be\label{index1}
CEI_t=\int_a^b |NC_T(s)-NC_N(s)|ds
\ee
is called the cohort effect index(CEI) of the generation born in year $t$. Here, for simplicity, we use the arc length $s$ as a parameter to describe the cohort curve $l_t$ and the integrating range $[a,b]$ is decided by the structure of mortality data.}\\

The reality is that the data of mortality are always discrete points on the surface $\Sigma_s$. We view the set of the points as a piecewise mortality surface and denote it by $\Sigma_p$ (Figure 2). Now we briefly describe how to realize the idea above on this piecewise surface $\Sigma_p$.

To begin our program, first we define the discrete parameter of a discrete curve $l$ contains three point $p_0$, $p_1$ and $p_2$. The discrete parameter is defined by
$$s_0=0,\ \ s_1=\frac{|p_1-p_0|}{|p_{1}-p_{0}|+|p_2-p_1|},\ \
s_2=1.$$
Next we estimate the tangent vector of $l$ at $p_1$. We call it a discrete tangent vector and denote it by $\vec{T}=(T_t, T_x, T_z)$. By minimizing the sum of the distances between the tangent line and the two points $p_0$ and $p_1$ under the constrain that the tangent line should pass through the point $p_1$, we can get an approximation of $\vec{T}$.
\be\label{tv1}
T_t=\frac{(s_0-s_1)\left(t(s_0)-t(s_1)\right)
+(s_2-s_1)\left(t(s_2)-t(s_1)\right)}{(s_0-s_1)^2+(s_2-s_1)^2},
\ee
\be\label{tv2}
T_x=\frac{(s_0-s_1)\left(x(s_0)-x(s_1)\right)
+(s_2-s_1)\left(x(s_2)-x(s_1)\right)}{(s_0-s_1)^2+(s_2-s_1)^2},
\ee
\be\label{tv3}
T_z=\frac{(s_0-s_1)\left(z(s_0)-z(s_1)\right)
+(s_2-s_1)\left(z(s_2)-z(s_1)\right)}{(s_0-s_1)^2+(s_2-s_1)^2}.
\ee

For a piecewise mortality surface $\Sigma_p$, We use $p_{ij}$ to denote points on the surface, where the subscript $i$ refer to year and the subscript $j$ refer to age and the coordinates of the point $p_{ij}$ in $\mathbb{R}^3$ are $(t_i,x_j,f(t_i,x_j))$. Since curvature of a curve $l_t$ at certain point $p_{ij}\in l_t$ is a local quality, to define the normal curvature on the discrete surface, it is appropriate to use only the points around $p_{ij}$ to compute the curvature. The minimal neighbourhood of $p_{ij}$ is consisting of the nine points $\{p_{i-1,j+1},p_{i,j+1},p_{i+1,j+1},p_{i-1,j},p_{ij},p_{i+1,j},
p_{i-1,j-1}, p_{i,j-1},p_{i+1,j+1}\}$. Among them, we view $l_1:\{p_{i-1,j-1},p_{ij},p_{i+1,j+1}\}$ and $l_2:\{p_{i-1,j+1},p_{ij},p_{i+1,j-1}\}$ as two discrete short curves whose tangent directions are taken as approximations of $T$ and $N$ in our definition of CEI.

By theories of differential geometry, for a smooth curve $l$ parameterized by $s$, suppose the unit tangent vector field along $l(s)$ is $\vec{V}(s)$,  then the curvature vector of the curve is defined by
\be\label{cv}
\vec{CV}(s)=\frac{\vec{V}'(s)}{|l'(s)|},
\ee
where $'$ is the derivative with respect to the parameter $s$. For the two discrete curve $l_1$ and $l_2$, by formulas (\ref{tv1}-\ref{tv3}) and normalization, we can get the unit discrete tangent vectors to $l_1$ and $l_2$ at point $p_{ij}$ and we denote them by $\vec{V}_1(p_{ij})=(v_{1t}(p_{ij}),v_{1x}(p_{ij}), v_{1z}(p_{ij}))$ and
$\vec{V}_2(p_{ij})=(v_{2t}(p_{ij}),v_{2x}(p_{ij}), v_{2z}(p_{ij}))$.
For a discrete curve, the derivative with respect to its discrete parameter can be defined by solving a similar constrained minimization problem as we do in estimating $\vec{T}$. Thus we can get the two discrete curvature vector fields $\vec{CV}_1$ and $\vec{CV}_2$ just following the formula (\ref{cv}).

Obviously, two unit tangent vectors $\vec{V}_1$ and $\vec{V}_2$ are not enough to determine a unique vector orthogonal to them. To get the normal vector of the surface $\Sigma_d$ at any point $p_{ij}$, we consider two more discrete curves across $p_{ij}$. Let $l_3:=\{p_{i-1,j},p_{ij},p_{i+1,j}\}$ and $l_4:=\{p_{i,j+1},p_{ij},p_{i+1,j-1}\}$, the same as what we have done for $l_1$ and $l_2$, we can get two unit tangent vectors $\vec{V}_3$ and $\vec{V}_4$. Since normal vectors are orthogonal to any tangent vector, we can estimate the discrete unit normal vector $\vec{N}(p_{ij})$ by minimizing
$$f(\vec{N})=\sum_{k=1}^{4} |\vec{N}\cdot \vec{V}_k|^2,$$
with the constraint $\vec{N}\cdot \vec{N}=1$. For details to solve the problem, one can refer to \cite{mitra}. Finally, it is nature to define the discrete normal curvature along direction $\vec{V}_k$ at point $p_{ij}$ by
$$NC_k(p_{ij})=N(p_{ij})\cdot \vec{CV}_k(p_{ij}), \ \ k=1,2,3,4$$

For a fixed integer $m$, all the points $p_{ij}$ satisfying $i+j=m$ make up a curve related to a group of people born in the same year. We call this group cohort $m$ or $C_{m}$ and call the curve a cohort curve. The tangent vector field along the cohort curve corresponds to $\vec{V}_1$ and we call this direction a cohort direction. By our definition of CEI for the smooth case, we define the discrete CEI for $C_{m}$ by
$$CEI_{m}=\sum_{i+j=m}|NC_1(p_{ij})-NC_2(p_{ij})|.$$
Now for any integer $m$ satisfying $a\leq m\leq b$, we get $CEI_m$. All these cohort effect indexes form a time series, and we call it the series of cohort effect (index) in the following of this paper.

\begin{figure}
\includegraphics[height=120pt,width=300pt]{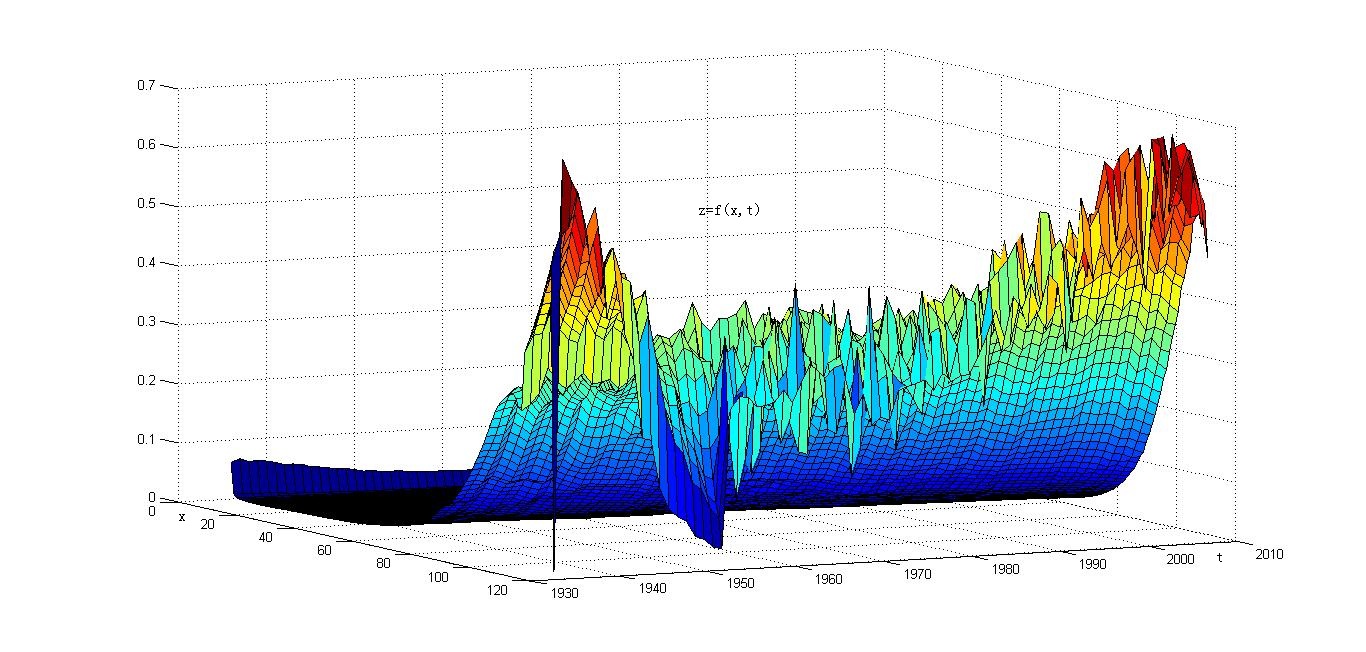}
\caption[surface]{An example of the mortality surface}
\end{figure}

\section{Applications of the approach to identify and measure cohort effects}

   The mortality data used in this paper are obtained from "The Human Mortality Database" \footnote{http://www.mortality.org/}. There are several types of data sets in this database. Since we want to get results as accurate as possible, we choose the data sets in which age and period are grouped with the shortest time interval (e.g. $1$ year). For auxiliary check or comparison, we also use the data sets with age interval $\times$ period interval equals to $1\times5$ and  $5\times 5$.

\begin{figure}[tbh]
\begin{centering}
\subfloat[first]{\begin{centering}
\includegraphics[height=90pt,width=185pt]{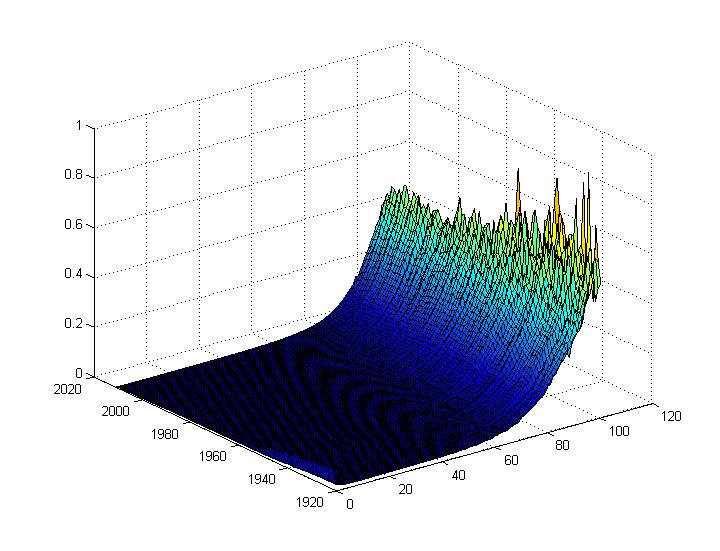}
\par \end{centering}
}
\subfloat[second]{\begin{centering}
\includegraphics[width=185pt]{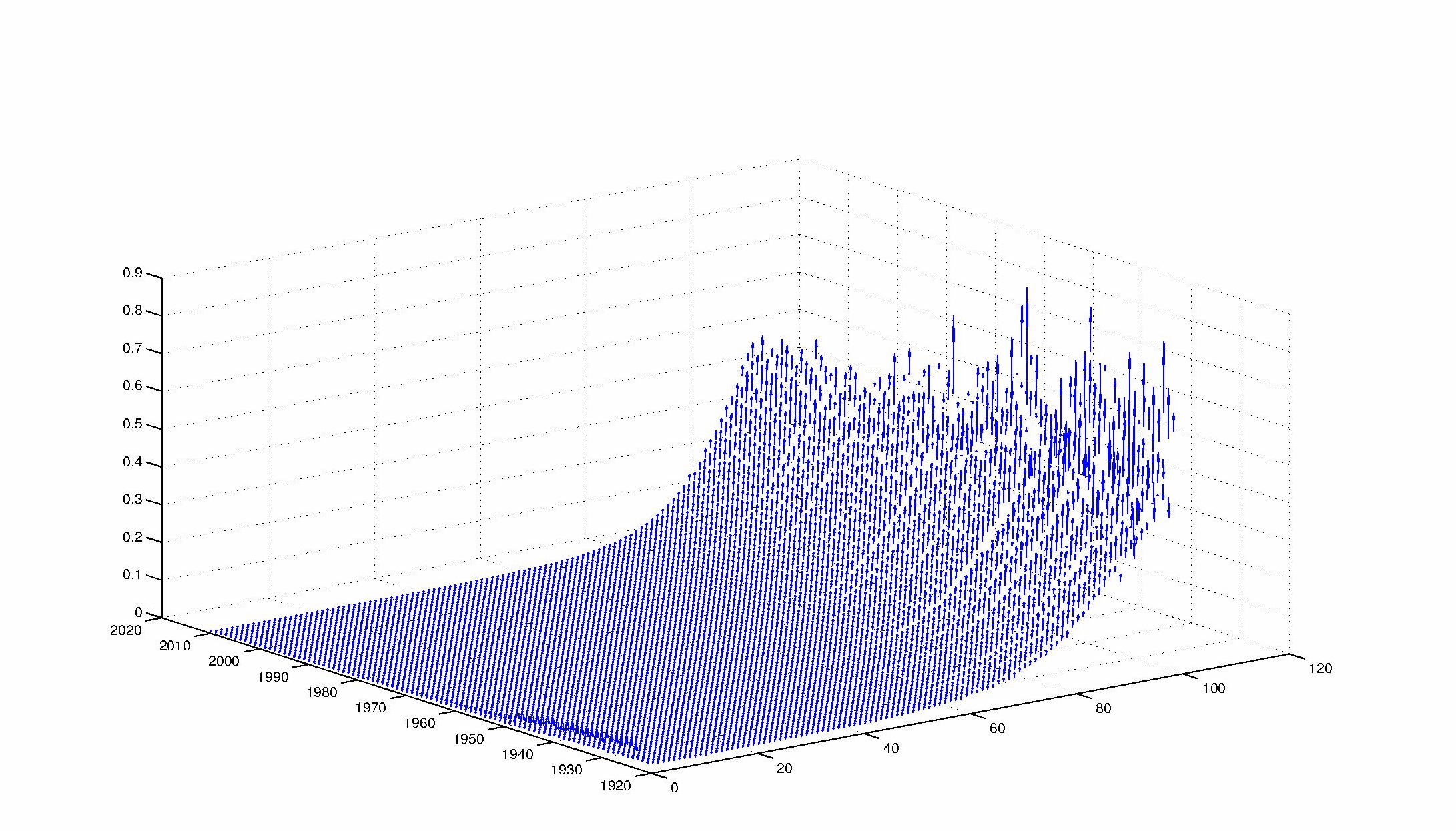}
\par\end{centering}
}

\subfloat[third]{\begin{centering}
\includegraphics[width=185pt]{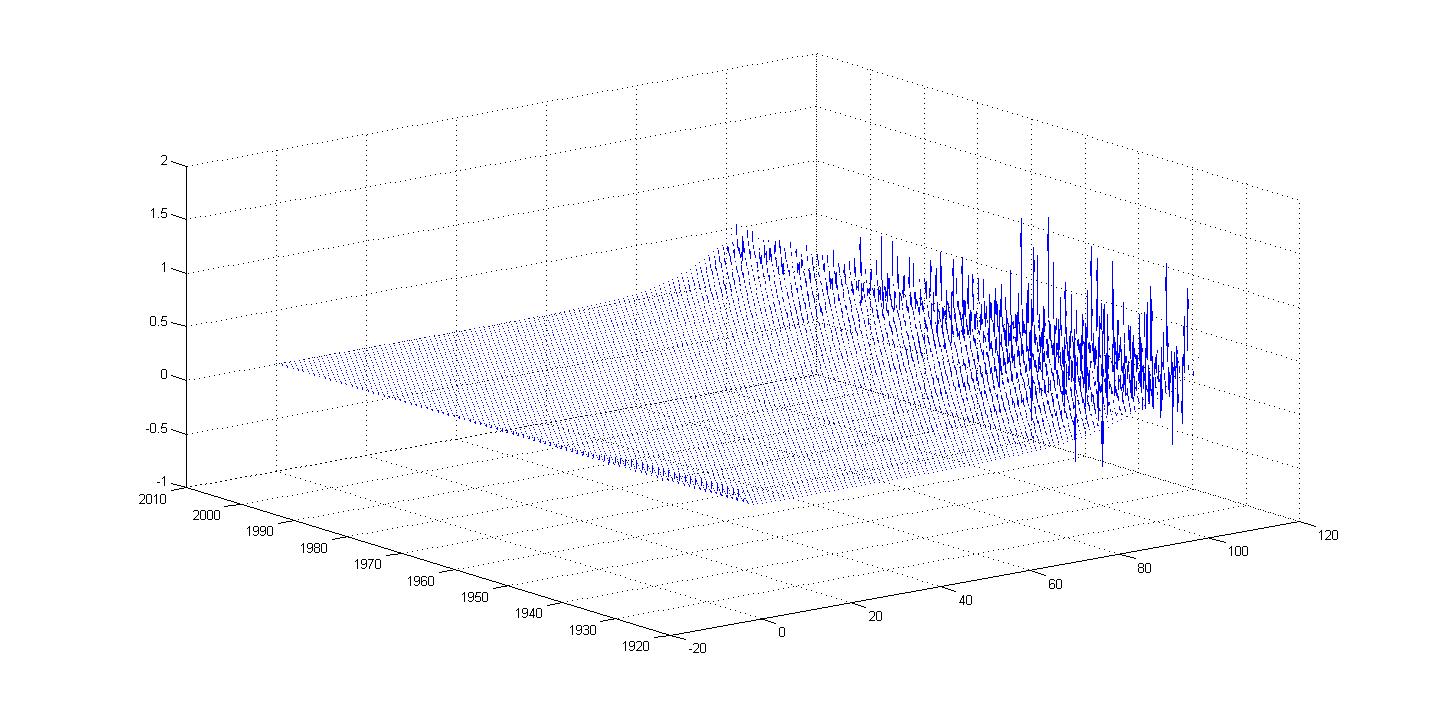}
\par\end{centering}
}
\subfloat[fourth]{\begin{centering}
\includegraphics[width=185pt]{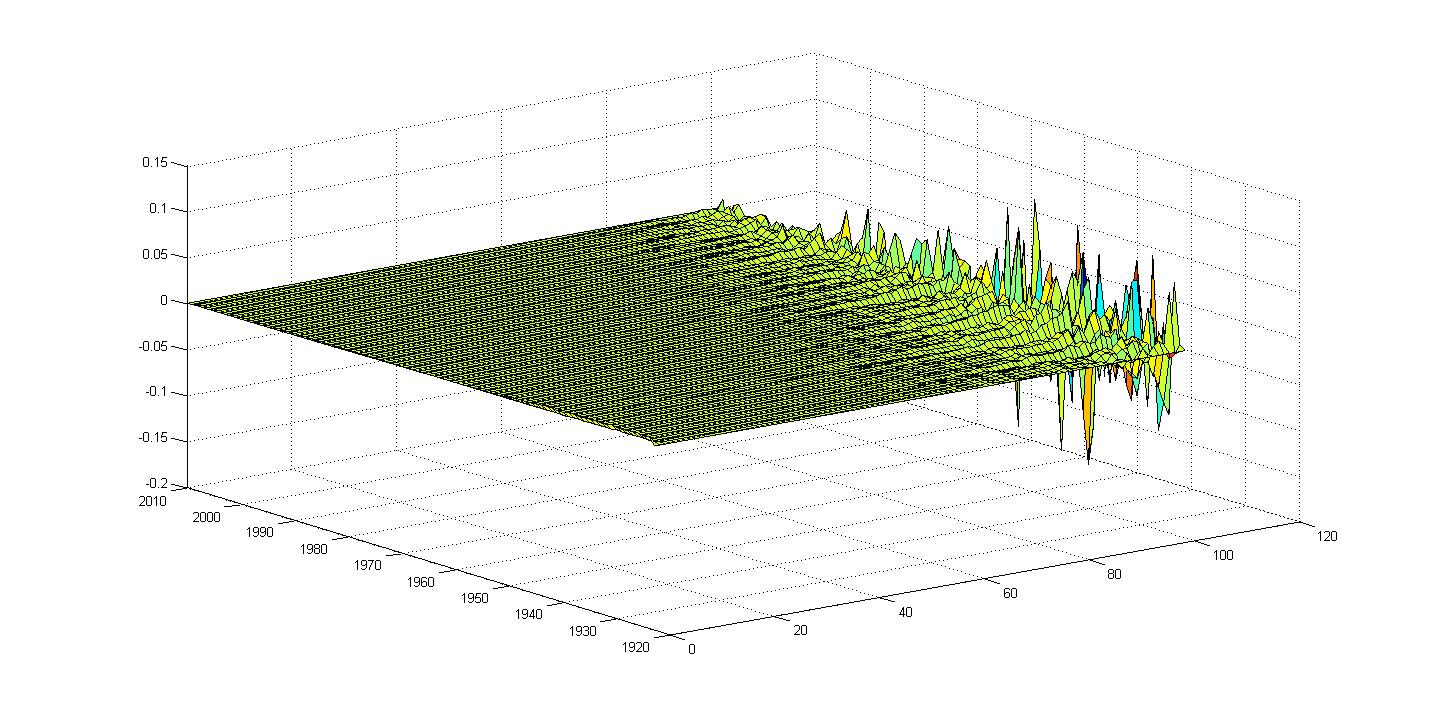}
\par\end{centering}
}

\par\end{centering}
\caption[first]{The process of calculating the series of cohort effect}
\end{figure}

\begin{figure}[tbh]
\begin{centering}
\subfloat[The cohort effect series of U.K.]{\begin{centering}
\includegraphics[height=90pt,width=185pt]{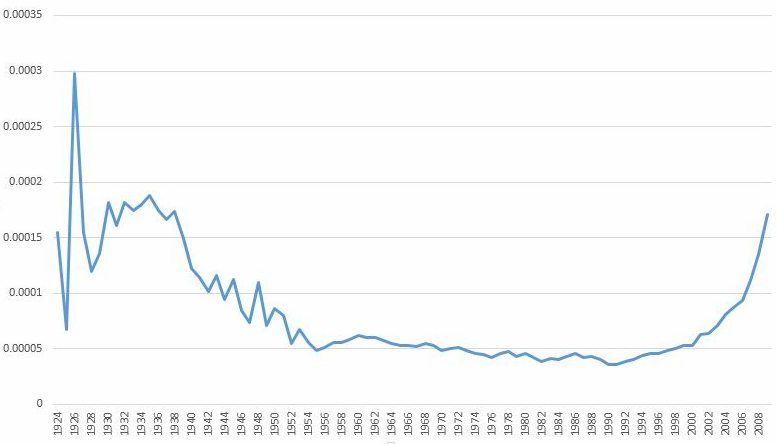}
\par \end{centering}
}
\subfloat[3-dimensional block graph]{\begin{centering}
\includegraphics[height=90pt,width=170pt]{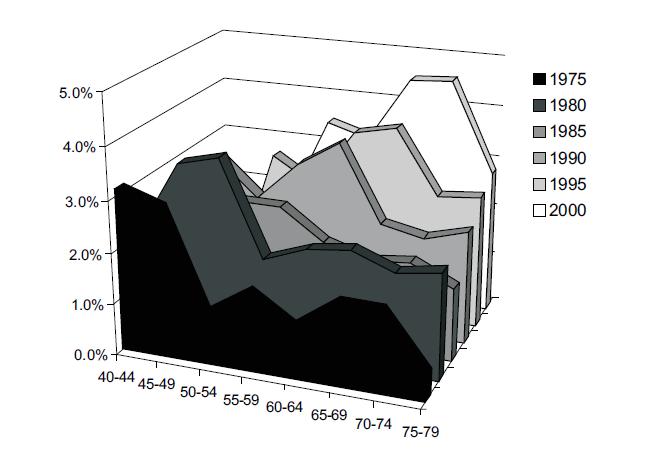}
\par \end{centering}
}
\end{centering}
\caption[uk]{The cohort effect of U.K.}
\end{figure}

  We first apply the model in Section 2 to the mortality data of U.K.. Its birth cohort effects are frequently referred to in references. The calculating process is shown  intuitively in Figure 3. The original mortality data set corresponds to a discrete mortality surface, shown by the upper left figure (a). The results of computations for tangent vectors and curvature vectors are four vector fields along different directions. The upper right figure (b) and the lower left figure (c) give the results along the cohort effect direction which is defined in Section 2. For brevity, we omit figures of other directions. Finally, the normal curvature along the direction of the cohort effect can be showed as another surface which is the lower right figure (d).

  Although we can get some direct impression about the mortality data from the surface of normal curvature, it is still difficult to recognize cohort effects exactly. Just as what has been described in Section 2, the detection can be carried out in the form of the cohort effect series. We draw the series on a plane. Then every peak of the series means that there exists obvious cohort effect since the mortality of this generation is apparently different from those of the neighbors.

  The left figure in Figure 4 gives the cohort effect series of U.K. which just comes up to our expectations. We can find two obvious peaks there. One peak happens on a interval of about $3\sim 5$ years and another peak lasts for about $10$ years which include several small peaks. The two peaks show strong birth-year cohort effects. This cohort effect series is in accord with the well-known U.K. cohort effect which has been noted several times in literatures. For example, the right figure in Figure 4 which comes from Willets \cite{willets} uses a three-dimensional block graph to show cohort effect of U.K. intuitively. The two graphs in Figure 4 both give the conclusion that the generations (of both sexes)who born approximately between 1925 and 1945 have experienced more rapid mortality changing. Moreover, more detailed information about cohort effects could be seen intuitively in our result ((a) in Figure 4):

   1)  By the theory of our model, the generation born in 1925 holds obvious mortality cohort effect since there is a highest peak in the final series. The reasons for that are not precisely understood. Maybe just data errors or outliers dominate it. For cohort effect detection, just one year is too short to represent one generation in view of scale.

   2) The generation born between 1930 and 1935 holds another obvious mortality cohort effect. The corresponding series is approximately a plateau but not a peak. $6$-year is enough to represent a generation and we view this generation as the dominating part of cohort effect mentioned in literatures. The reason that we can reduce the gap in literatures from $[1925,1945]$ to $[1930-1935]$ is nothing but calculation.

\begin{figure}[tbh]
\begin{centering}
\subfloat[U.K.]{\begin{centering}
\includegraphics[height=80pt,width=185pt]{ukfig.jpg}
\par \end{centering}
}
\subfloat[U.S.]{\begin{centering}
\includegraphics[width=185pt]{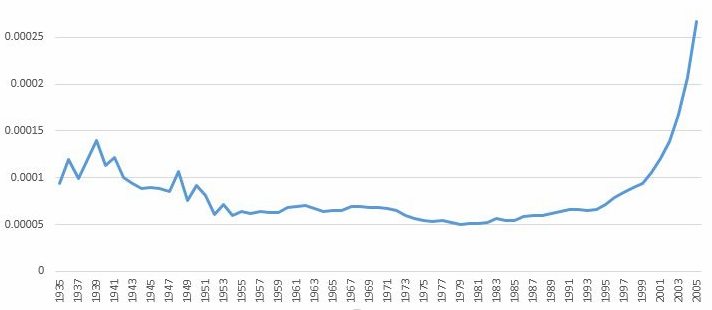}
\par \end{centering}
}

\subfloat[Canada]{\begin{centering}
\includegraphics[width=185pt]{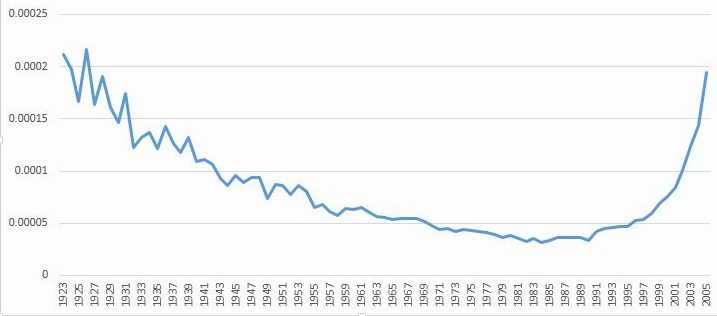}
\par \end{centering}
}
\subfloat[Japan]{\begin{centering}
\includegraphics[width=185pt]{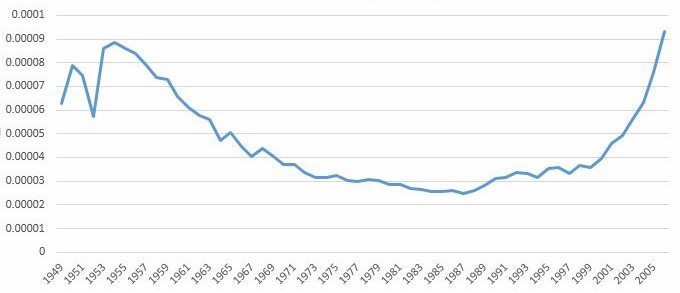}
\par \end{centering}
}
\end{centering}
\caption[uk]{Series of cohort effect }
\end{figure}

     We also apply our method to the data sets of other three countries and the series of cohort effects are shown in Figure 5. The upper left one is the series of U.K., the upper right one is the series of U.S, the lower left one is the series of Canada and the lower right one is the series of Japan. By discussions similar to what we have done for U.K., we assert that, for the series of U.S., cohort effects exist in the gap [1933,1941] and the gap [1948,1953] and there is no other obvious cohort effect. The situation similar to U.S. exists in the series of Japan. But the gaps are [1949,1952] and [1953,1958]. The volatility of the series of Canada is more drastic and this means that mortality of neighboring generations are much more different.

    There is one common ground that all of the four series have a stable upward trend and then a sharp drop at the end. Here we provide an explanation for this kind of behavior of the series.

    1) The upward trend is the consequence of shorter data. There is fewer and fewer data along the same cohort direction, for example, only one data in the last birth year (with regard to age 0 and calender year 2010). This kind of data cannot reflect the global situation of the corresponding cohort and leads to much volatility in calculating normal vectors based on our discrete algorithm of differential geometry. So the oscillations of mortality for the last few generations are bigger than those for the earlier generations. This definitely brings errors for measuring cohort effects and we should just consider the former part of the series of cohort effect and remove the tail of the youngest generations. In fact, it is unpersuasive to claim any cohort effect for a group when we can only capture little information about their mortality. We recommend to detect and measure cohort effects for the generations who born before at least 1970.

    2) The cause of the sharp drop in the tail of the series is because of our initial conditions for differential calculations. In fact, we set the boundary data to be $0$ so that the size of the data set remains the same as the size of the original data set in the process of differential calculations.

    As what we have just explained, the sharp drops of the series are useless when we deal with cohort effects, so we cut off the tails of the series in our figures of series of cohort effect.

    We also find that "U" shape is a general result when we apply our method on the data sets of more countries. The bottom of "U" shape means that the generations who born between 1950 and 1980 experience stable mortality improvement. Furthermore, the mortality volatility tends to decrease since 1940 in most of countries. It is convinced that the post-war prosperity of most countries contributes to "U" shape.

\begin{figure}[tbh]
\begin{centering}
\includegraphics[height=120pt,width=300pt]{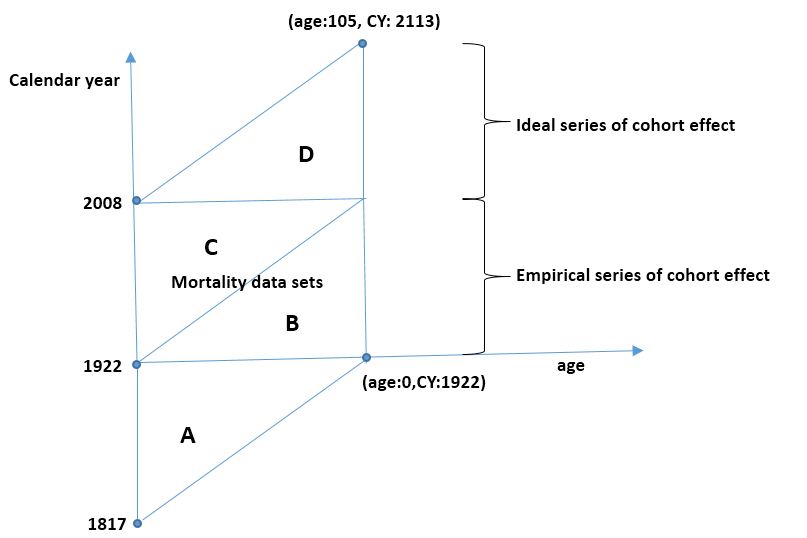}
\end{centering}
\caption[dataset]{The data sets and series of cohort effect}
\end{figure}

\begin{figure}[tbh]
\begin{centering}
\includegraphics[height=100pt,width=300pt]{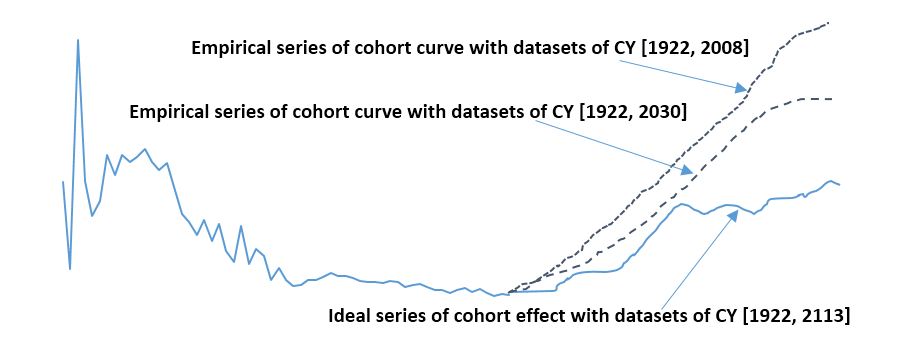}
\end{centering}
\caption[dataset]{Ideal and empirical series of cohort effect}
\end{figure}

     "U" shape is also related to the length of the data sets. Here we make a formal description with the auxiliary help of Figure 6 and Figure 7. If the data sets are complete, we can get an ideal series of cohort effect by using all need data which includes Part A, Part B, Part C and Part D (Figure 6).  Taking U.K. as an example, if we track the generations from 1922 to 2008, the mortality at age 105 in calendar year 2113 is needed (set the maximal age to be 105). But this is impossible. Usually a mortality data set is a data matrix which consists of Part B and  Part C, as shown in Figure 6. Since we cannot track all the targeted generations from their birth to death, we only work on Part C of the data set. By this way, we get an empirical series of cohort effect instead of an ideal series of cohort effect. So the above series of cohort effect in Figure 5 (U.K., U.S., Canada and Japan) are actually the empirical series of cohort effect. The series which is calculated from part C and D of the data set are called the ideal series of cohort effect. There is a direct relationship between ideal and empirical ones. When a new row of data (for example, the mortality data for all ages in 2009) is added into the data set of Part C, the new empirical series of cohort effect will be closer to the ideal series of cohort effect. The ideal series of cohort effect shows the accurate information about cohort effect but it is impossible to obtain. Since the empirical series of cohort effect is a good approximation of the ideal series of cohort effect especially for early generations, we can use the empirical series of cohort effect to find desired information, except for recently born generations. .

     From the above analysis, we know that the left part of 'U' can provide us desired information about cohort effect and the right part of 'U' does not provide any believable information because of the insufficiency of the data set.

\section{Applications to mortality models}

The Lee-Carter and APC models are two widely used mortality models in longevity risk. The Lee-Carter model considers only age and period effects. To improve the model, several APC models which take cohort effects into consideration are developed. But they all need additional conditions on structures of parameters in the models. Now having the quantitative estimates of cohort effects, we can try to apply the estimates in the APC model.

Firstly, we subtract a rescaled $CEI_m$ for each cohort to remove the corresponding cohort effect and polish the mortality surface. In this step, we should use a unique scale to rescale $CEI_m$ for each data set. Since our estimates are dimensionless and the relative relations of cohort effects for different groups in the same data set remain unchanged, what we have done is reasonable. After this kind of pretreatment of the data set, we can deal with the new data set by using the Lee-Carter model. We call this approach a modified Lee-Carter (MLC) model.

We apply the classical Lee-Carter model to the mortality data of U.K., U.S., Canada and Japan first. Then we analyze the same data sets by using the MLC model. The results for $k_t$ and $\beta_x$ are shown in Figure 8 and 9. The dotted lines by $*$ represent results by the Lee-Carter model and the full lines represent results by the MLC model. In these two figures, results by the Lee-Carter model and by the MLC model are similar to each other.  Our pretreatment of the data sets works well and there is no obvious distinctions of both $k$ and $\beta$ values in these two models.

\begin{figure}[tbh]
\begin{centering}
\subfloat[U.K. k-value]{\begin{centering}
\includegraphics[height=115pt,width=185pt]{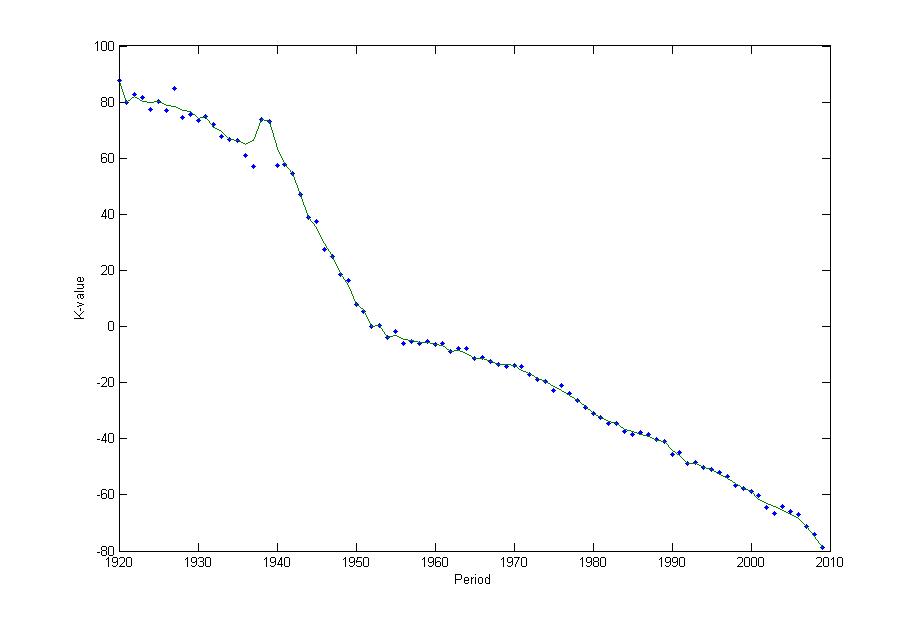}
\par \end{centering}
}
\subfloat[U.S. k-value]{\begin{centering}
\includegraphics[width=185pt]{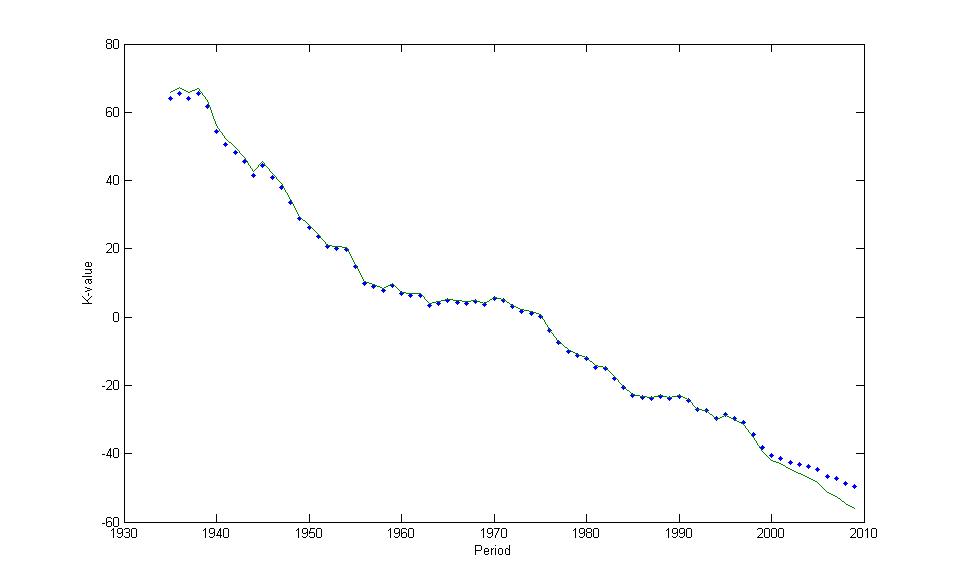}
\par \end{centering}
}

\subfloat[Canada k-value]{\begin{centering}
\includegraphics[height=120pt,width=185pt]{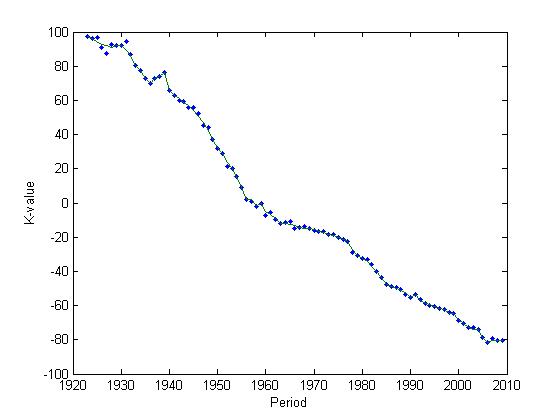}
\par \end{centering}
}
\subfloat[Japan k-value]{\begin{centering}
\includegraphics[height=120pt,width=185pt]{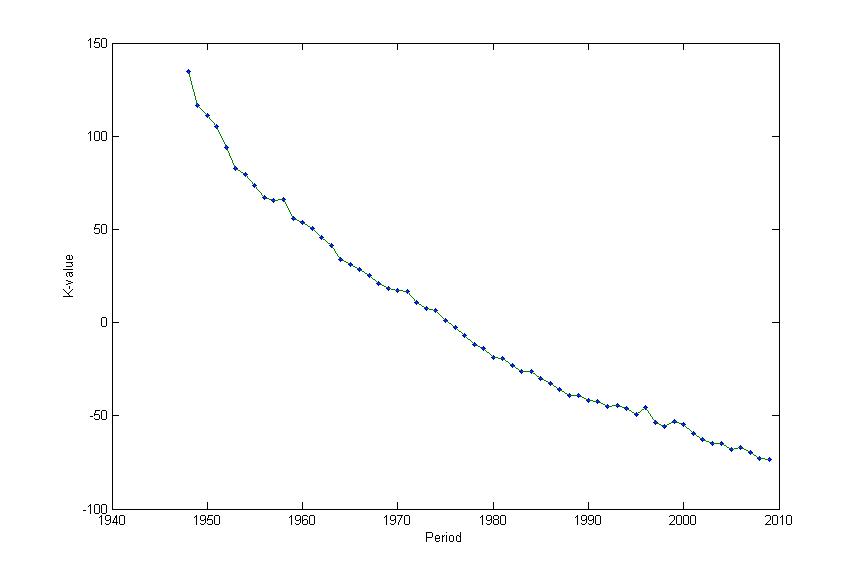}
\par \end{centering}
}
\end{centering}
\caption[uk]{$k$-value}
\end{figure}

\begin{figure}[tbh]
\begin{centering}
\subfloat[U.K. $\beta$-value]{\begin{centering}
\includegraphics[height=115pt,width=185pt]{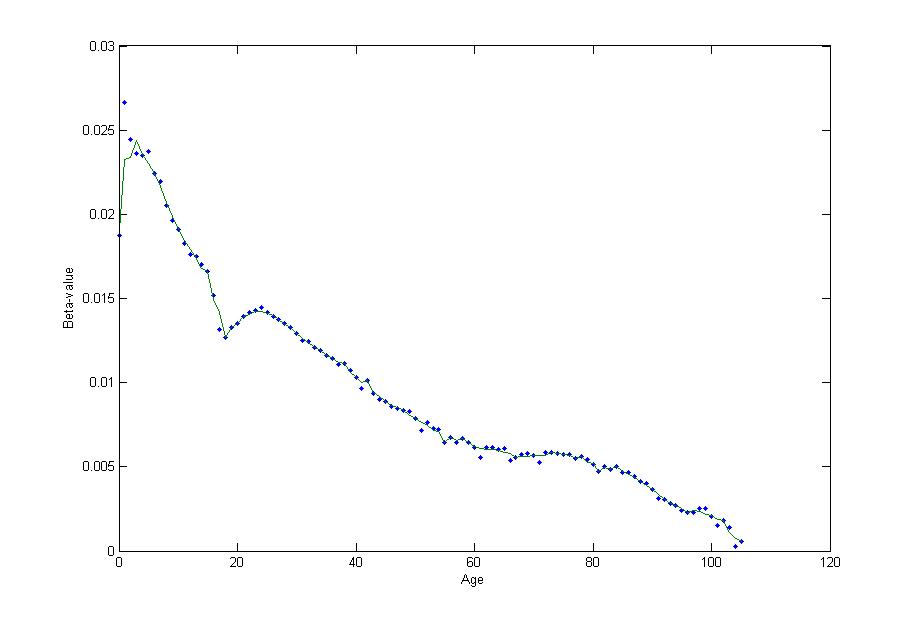}
\par \end{centering}
}
\subfloat[U.S. $\beta$-value]{\begin{centering}
\includegraphics[width=185pt]{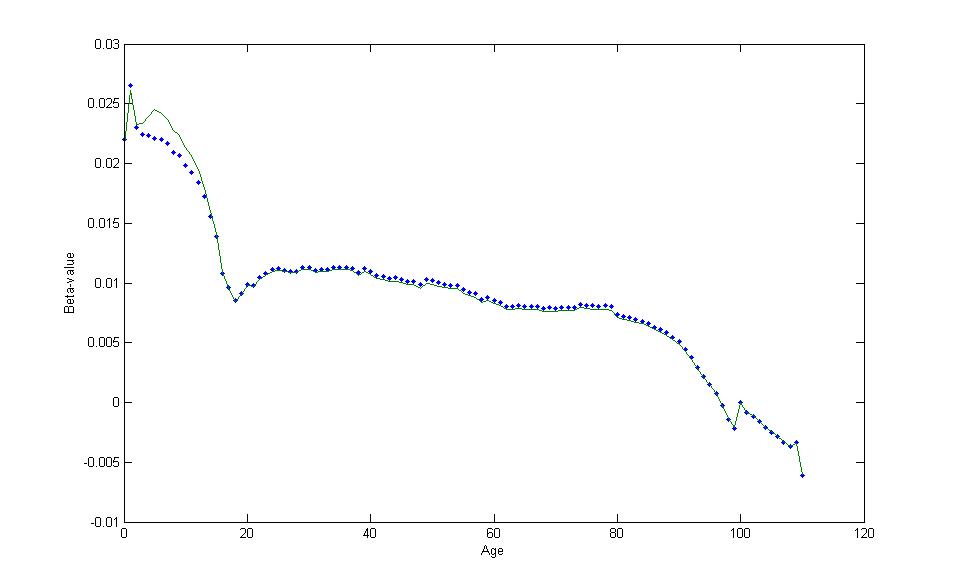}
\par \end{centering}
}

\subfloat[Canada $\beta$-value]{\begin{centering}
\includegraphics[height=120pt,width=185pt]{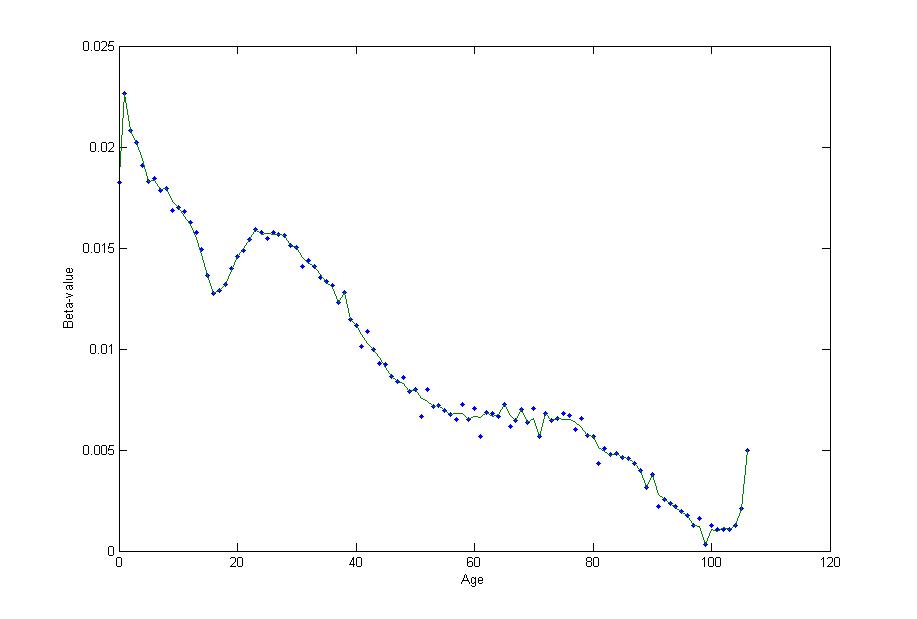}
\par \end{centering}
}
\subfloat[Japan $\beta$-value]{\begin{centering}
\includegraphics[height=120pt,width=185pt]{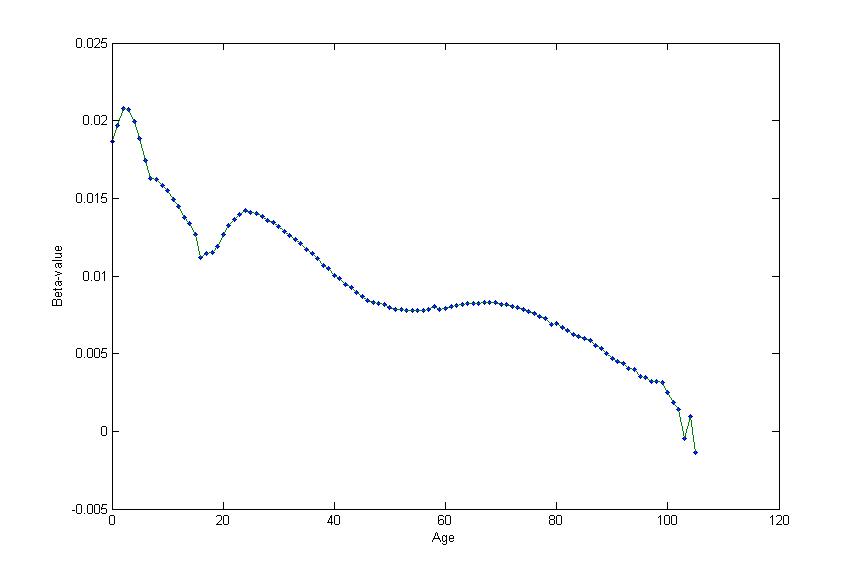}
\par \end{centering}
}
\end{centering}
\caption[uk]{$\beta$-value}
\end{figure}

We next try to use the MLC model and the APC model to forecast mortality for U.S.. When we apply the MLC model, we use the algorithm for the Lee-Carter model after pretreating the data sets. When apply the APC model, we follow the algorithms in \cite{brouhns,pitacco,renshaw2,wilmoth}. The results are shown in Figure 10. Here the dotted lines by $*$ are predicted values of mortality by using the MLC model. The dotted lines by $\cdot$ and $+$ are predicted values of mortality by using the APC model with two different algorithms. The full lines are actual data of mortality in 2010 and 2011.

Since the mortality for young people and old people (age less than $20$ and older than $50$) are unstable and contain more noises, we only intercept the results for age in the interval $[20, 50]$. For data of U.K. and Canada, the results are similar and we omit them here for brevity. For data of Japan, we find that our approach is not as good as the APC model. The reason should be that the size of the data set of Japan is smaller than other three. Since cohort effect is a global property, less data should cause more deviation when we estimate cohort effect. For this reason, we recommend to use our approach for data sets with age interval bigger than $60$ years.

\begin{figure}[tbh]
\begin{centering}
\subfloat[Prediction of 2010]{\begin{centering}
\includegraphics[height=140pt,width=185pt]{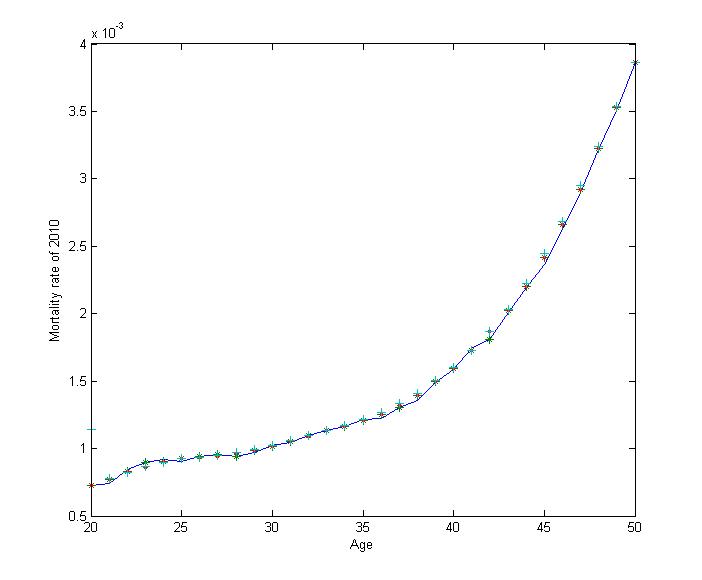}
\par \end{centering}
}
\subfloat[Prediction of 2011]{\begin{centering}
\includegraphics[height=140pt,width=185pt]{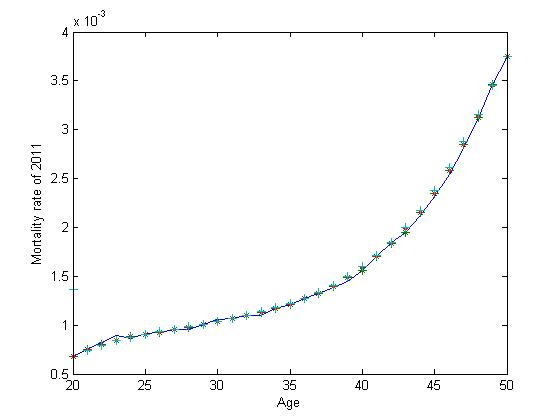}
\par \end{centering}
}
\end{centering}
\caption[prediction]{Prediction of mortality}
\end{figure}

For the predicted results of U.S. in Figure 10, it seems that there is still no obvious distinction when applying different models. To make it more clear, in Figure 11, we intercept a part of the results and exaggerate the curves. We find that sometimes, our results are closer to the actual data.

\begin{figure}[tbh]
\begin{centering}
\includegraphics[height=120pt,width=200pt]{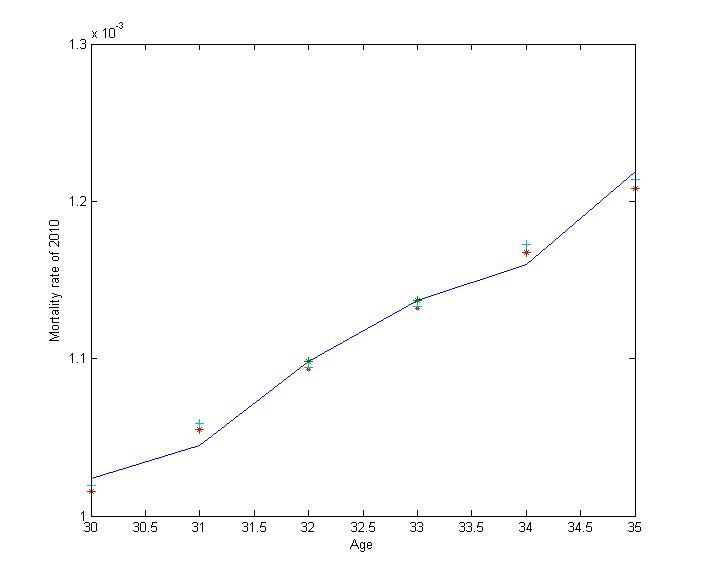}
\end{centering}
\caption[Predict]{Predicted mortality of 2010}
\end{figure}

\section{Conclusion and the future work}

   We promote an effective method based on differential geometry to implement quantitative measurement of cohort effect. The peaks on the series of cohort effect mean the existence of cohort effects and the height of the peaks tells us the strength of the corresponding effects. We also apply this method on the data sets of four countries including the United Kingdom, the United states, Canada and Japan. All the resulting series show the desired strength of cohort effects in different generations. In particular for U.K., our method can give a further description of the well-known mortality cohort effect.

   Applications of our estimates in the analysis on longevity risk are one of the problems we are considering. We develop a modified Lee-Carter model and our estimates works well in this model. We also give predicted values of mortality for U.S.. There is no obvious distinction for results by using our model and the APC model and sometimes our approach seems better. But the underlying causes of the difference can only be confirmed after research work of the mathematical structure and statistical analysis of enough samples. This should be done in future work. Besides, further studies on the theoretical model are in progress. We hope to make the model subtler and more intelligent and analyze the contributing factors of the detected cohort effects.

   Applications of the estimates on cohort effect of different types of data set, such as mortality of some epidemics, is also an interesting topic. We hope that this kind of work could give some guidance to find the underly causes of diseases and to make the health policy.

{\bf Acknowledgement.}
 The first author is partially supported by the Funding of Beijing Philosophy and Social Science (No.15JGC153), the MOE Project of Key Research Institute of Humanities and Social Sciences at Universities (No.11JJD790004) and Program for Innovation Research in Central University of Finance and Economics. The second author is partially supported by NSFC 11201028 and the Fundamental Research Funds for the Central Universities. Both authors thank the support of Data Lighthouse Plan.

\end{document}